\documentclass[conference,letterpaper]{IEEEtran}
\usepackage{graphicx}
\usepackage{epsfig}
\usepackage{epstopdf}
\usepackage{amsmath,amsthm,amsfonts,amssymb}
\usepackage{multirow}
\usepackage{cite}
\usepackage{tikz}
\usepackage{ellipsis}
\usepackage{xcolor}
\usepackage{xifthen}
\usepackage{cancel}
\usepackage{amsmath,bm}
\usepackage{balance}
\usepackage{array}
\usepackage[linesnumbered, ruled, vlined, noend]{algorithm2e}
\theoremstyle{definition} 
\theoremstyle{definition} 
\theoremstyle{definition} 
\theoremstyle{definition} \newtheorem{cor}{Corollary}
\newtheorem{theorem}{Theorem}

\newcommand{\GA}{\text{GA}}
\newcommand{\UTL}{\text{UTL}}
\newcommand{\RM}{\text{RM}}
\newcommand{\Aut}{\text{Aut}}
\newcommand{\Aff}{\text{Aff}}
\newcommand{\fixme}[2]{\ifx&#2&{\leavevmode\color{red}#1}\else{\leavevmode\color{red}FIXME\{}#1{\leavevmode\color{red}\}}\footnote{{\leavevmode\color{red}#2}}\PackageWarning{Fixme}{#1: #2}\fi}
\newcolumntype{M}[1]{>{\centering\arraybackslash}m{#1}}

\begin{document}

\title{Polar Codes for Automorphism Ensemble Decoding}

\author{\IEEEauthorblockN{Charles Pillet, Valerio Bioglio, Ingmar Land}
\IEEEauthorblockA{Mathematical and Algorithmic Sciences Lab\\ Paris Research Center, Huawei Technologies Co. Ltd.\\
Email: $\{$charles.pillet1,valerio.bioglio,ingmar.land$\}$@huawei.com}} 

\maketitle

\begin{abstract}
%
In this paper we deal with polar code automorphisms that are beneficial under low-latency automorphism ensemble (AE) decoding, and we propose polar code designs that have such automorphisms.  
Successive-cancellation (SC) decoding and thus SC-based AE decoding are invariant with respect to the only known polar code automorphisms, namely those of the lower-triangular affine (LTA) group.  
To overcome this problem, we provide methods to determine whether a given polar code has non-LTA automorphisms and to identify such automorphisms.  
Building on this, we design specific polar codes that admit automorphisms in the upper-diagonal linear (UTL) group, and thus render SC-based AE decoding effective.   
Demonstrated by examples, these new polar codes under AE decoding outperform conventional polar codes under SC list decoding  in terms of error rate, while keeping the latency comparable to SC decoding. 
Moreover, state-of-the-art BP-based permutation decoding for polar codes is beaten by BP-based AE thanks to this design.
\end{abstract}


\section{Introduction}
\label{sec:intro}
Polar codes \cite{ArikanFirst} are a class linear block codes relying on the phenomenon of channel polarization.  
They are shown to be capacity-achieving on binary memoryless symmetric channels under successive cancellation (SC) decoding for infinite block length.  
In the finite-length regime, however, SC decoding suffers from poor performance due to error propagation.  
In SC list (SCL) decoding, multiple paths are followed in parallel SC decoders, and the best path is selected from the list at the end \cite{TalSCL}.  
In connection with an outer CRC, SCL decoding provides excellent error rate results \cite{CRC_aid}, making this SCL-CRC decoding the de-facto standard and often used performance benchmark for polar code decoding.
%
However, SCL decoders need to regularly pause decoding operations to exchange information, increasing the decoding delay.  
This may be avoided by permutation decoders, where several decoders run independently in parallel on permuted codewords. 
To this end, permutations based on stage permutations of the polar codes were exploited: this method is analyzed under SC decoding in \cite{PermGross}, while soft cancellation (SCAN) \cite{SCANfirst} is applied in \cite{SCANL} and belief propagation (BP) \cite{BPfirst} is chosen in \cite{BPLRM}. 
\begin{figure}
	\centering
	\includegraphics[width=0.45\textwidth]{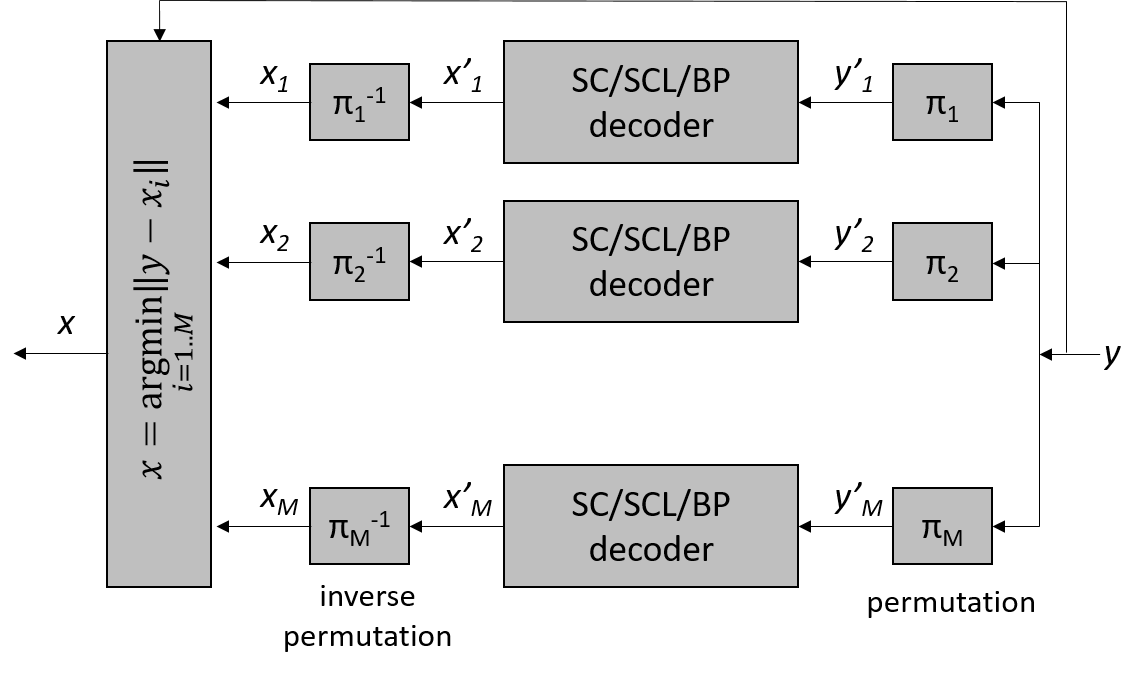}
	\caption{Structure of the automorphism ensemble (AE) decoder.}
	\label{fig:AE}
\end{figure} 

Recently, a new type of permutation group decoder for Reed-Muller codes \cite{RMcode,RMcodeREED}, termed \emph{automorphism ensemble} (AE) decoder, has been proposed based on the automorphism group of the code \cite{geiselhart2020automorphism}.
The automorphism group of Reed-Muller codes is known and equivalent to the general affine group \cite{BardetPolyPC}, which allows a large set of automorphisms for AE decoding.
AE decoding may be applied to polar codes given the similarity of the two codes; however it requires knowledge of polar code automorphisms.
Lower-triangular affine (LTA) transformations being a known subgroup of the automorphism group of polar codes \cite{BardetPolyPC}, researchers used this group under AE decoding; however, SC decoding was proved to be invariant under LTA automorphisms, and SC-based AE decoding exhibits no gain using these permutations \cite{geiselhart2020automorphism}. 

This paper deals with polar-code automorphisms that lie outside the LTA group.  
We propose a method to determine if a given polar code has such automorphisms and to identify them. 
We further propose a polar code design that aims at constructing polar codes havinge sufficiently many non-LTA automorphisms to allow for efficient AE decoding. 
Such designed code can be classified as partially symmetric monomial code \cite{Urbanke_symmetry}.
Very recently, a new subgroup of automorphisms group of polar codes have been identified in the block lower triangular (BLTA) group \cite{geiselhart2021automorphismPC}.  
Similarly, our paper focus on upper-diagonal linear (UTL) transformations, since they are not invariant with respect to SC decoding. 
The use of UTL and LTA automorphism under BP is also novel for polar codes.
By error-rate simulations, we show that newly designed polar codes under AE decoding can outperform conventional polar codes under SCL decoding in terms of error rate, while the AE decoding latency is only in the range of its inner decoder i.e. BP or SC decoding.

\section{Preliminaries}
\label{sec:pre}
\subsection{Polar Codes}

A $(N,K)$ polar code of length $N=2^n$ and dimension $K$ is a binary block code defined by the kernel matrix $T_2\triangleq [ \begin{smallmatrix} 1 & 0\\ 1 & 1 \end{smallmatrix} ]$, the transformation matrix $T_N = T_2^{\otimes n}$, an information set $\mathcal{I} \in [N]$ and a frozen set $\mathcal{F} = [N] \backslash \mathcal{I}$, $[N] = \{0,1,\ldots,N-1\}$.  
For encoding, an auxiliary input vector $\boldsymbol{u}  = ( u_0,u_1,\dots, u_{N-1} )$ is generated by assigning $u_i = 0$ for $i \in \mathcal{F}$ (frozen bits), and storing information in the remaining entries, $i \in \mathcal{I}$.  
The codeword is then computed as $\boldsymbol{x} = \boldsymbol{u} \cdot T_N$.
The information and frozen set are typically selected according to the reliabilities of the virtual bit-channels resulting from the polarisation. 
Reliabilities can be determined through different methods \cite{frozenset}.

Successive Cancellation (SC) decoding is the fundamental decoding algorithm for polar codes and is proved to be capacity-achieving at infinite block length \cite{ArikanFirst}.
SC list (SCL) decoding was proposed to improve the polar code performance for finite code lengths \cite{TalSCL}. 
The concatenation of a CRC code \cite{CRC_aid}, used as a genie, provides excellent performance and SCL-CRC is currently the best decoding algorithm of polar codes.

BP decoding is a popular message passing decoder conceived for codes defined on graphs. 
This soft-input/soft-output decoder has been adapted to polar codes in \cite{BPfirst}. 
BP is parameterized by its number of iterations over the factor graph $I$.

\subsection{Monomials codes}
\label{subsec:monomial_code}

Monomial codes of length $N=2^n$ are a family of codes that can be obtained as evaluations of boolean functions, namely as polynomials in $\mathbb{F}_2[x_0,\dots,x_{n-1}]$. 
Polar codes and Reed-Muller codes can be described through this formalism \cite{BardetPolyPC}. 
In fact, the rows of $T_N$ represent all possible evaluations of negative monomials over $\mathbb{F}_2^n$ \cite{BardetPolyPC}, where a negative boolean variable $\bar{x}_i$ is given by $\bar{x}_i = \neg x_i = (1 \oplus x_i)$.
This can be proven by induction on $n$ given that the two rows of $T_2$ are the evaluation of two monomials over $\mathbb{F}_2$, namely the constant monomial $1$, evaluating to $(1,1)$, and the monomial $\overline{x}_0$, evaluating to $(1,0)$.
Table~\ref{tab:monomials} lists all different negative monomials over $\mathbb{F}_2^3$, their degrees and their evaluations.

A monomial code of length $N=2^n$ and dimension $K$ is generated by $K$ (positive or negative) monomials out of the $N$ monomials over $\mathbb{F}_2^n$. 
These $K$ chosen monomials form the \emph{generating monomial set} $\mathcal{M}$ of the code, while their evaluations, along with the evaluations of their linear combinations, provide the codebook of the code. 
A monomial code is called \emph{decreasing} if $\mathcal{M}$ includes all factors of every monomial in the set and their sub-factors \cite{BardetPolyPC}. 



\begin{table}[htb]
	\caption{Negative monomials for $n=3$ and their evaluations.}
	\label{tab:monomials}
	\centering
	\begin{tabular}{c|c|c|rc}
		Degree               & Monomial                      & Evaluation & \multicolumn{2}{r}{Index of row in $T_N$}  \\
		\hline
		$0$                  & $1$                           & $11111111$ & \quad 7      &  111 \\
		\hline
		\multirow{3}{*}{$1$} & $\bar{x}_0$                   & $10101010$ & 6            &  110 \\
		                     & $\bar{x}_1$                   & $11001100$ & 5            &  101 \\
		                     & $\bar{x}_2$                   & $11110000$ & 3            &  011 \\
		\hline
		\multirow{3}{*}{$2$} & $\bar{x}_0\bar{x}_1$          & $10001000$ & 4            &  100 \\
		                     & $\bar{x}_0\bar{x}_2$          & $10100000$ & 2            &  010 \\
		                     & $\bar{x}_1\bar{x}_2$          & $11000000$ & 1            &  001 \\
		\hline
		$3$                  & $\bar{x}_0\bar{x}_1\bar{x}_2$ & $10000000$ & 0            &  000
	\end{tabular}
\end{table}

Reed-Muller codes, which may be seen as polar codes with particular frozen sets, are monomial codes generated by all monomials up to a certain degree. 
Polar codes select generating monomials following the polarisation effect; if the polar code design is compliant with universal partial order, the resulting code is provably decreasing monomial \cite{BardetPolyPC}.

\subsection{Code Automorphisms}
\label{subsec:autocode}

An automorphism $\pi$ of a code $\mathcal{C}$ is a permutation that maps any codeword $x \in \mathcal{C}$ into another codeword $x' \in \mathcal{C}$. 
The automorphism group $\Aut(\mathcal{C})$ of a code $\mathcal{C}$ is the set containing all automorphisms of code $\mathcal{C}$.
It is well known that the automorphism group of Reed-Muller codes of length $N=2^n$ is given by the affine transformation group of order $n$, $\GA(n)$. 
This group is defined as the set of all transformations of $n$ variables described by  
\begin{align}
	x & \mapsto x' = A x + b 
	\label{eq:GA}
\end{align}
$x , x' \in \mathbb{F}_2^n$, where $A$ is an $n\times n$ binary invertible matrix and $b$ is a binary column vector of length $n$.

Though the automorphism group of polar codes is unknown, it is proved in \cite{BardetPolyPC} that a subgroup of the automorphisms of decreasing monomials codes is given by the lower-triangular affine (LTA) group $\text{LTA}(n)$. 
This is the sub-group of $\text{GA}(n)$ for which $A$ is lower-triangular with ones on the diagonal. 
Since well-designed polar codes are typically decreasing monomials codes, this leads to the knowledge of one automorphism subgroup of polar codes.

\subsection{Automorphism ensemble decoding}
\label{subsec:AE}

In \cite{geiselhart2020automorphism}, the authors proposed an automorphism ensemble (AE) decoder for Reed-Muller codes, as depicted in Figure~\ref{fig:AE}:
$M$ SC-based decoders are run in parallel, each one starting from a permuted version of the received codeword where each permutation belongs to the automorphism group of the code. 
The codewords resulting from each SC unit are permuted back, and the most likely candidate is selected.  
%
Even though the structure of this decoder is similar to the one of SCL, the main difference is given by the resulting decoding latency: since the SC-based decoders do not need to exchange information during the decoding process, the latency of AE is essentially given by the latency of a single SC decoder.  
The authors of \cite{geiselhart2020automorphism} use BP, SC and SCL algorithms as inner decoders for AE, resulting in different error rate performances and latencies.

\section{Automorphisms of polar codes}
\label{sec:auto_PC}
In the following, using the concept of monomial codes, we analyze and provide new automorphism of polar codes.
As proved in \cite{geiselhart2020automorphism}, automorphisms within the LTA sub-group are absorbed by SC decoding.
Since permutations in LTA are the only automorphisms known for polar codes, in order to exploit AE decoders, we need to find new automorphism not in the LTA sub-group. 
In the following, we will restrict this search to upper-triangular linear (UTL) transformations, which are transformations of $n$ variables described by 
\begin{equation}
	x \mapsto x' = U x
	\label{eq:UTL}
\end{equation} 
$x , x' \in \mathbb{F}_2^n$, where $U$ is an $n\times n$ upper-diagonal binary matrix with full diagonal. 
This choice is motivated by our observation that UTL automorphisms are typically not absorbed by SC decoding and thus useful candidates for SC-based AE decoding.

\begin{table}[bt]
	\caption{Variable change via binary extension.}
	\label{tab:variable-change}
	\centering
	\scriptsize
	\setlength{\tabcolsep}{3pt}
	\begin{tabular}{|r|c|c|c|c|r|}
		\cline{2-6}
		\multicolumn{1}{c|}{} & \multicolumn{5}{c|}{index}\\
		\cline{2-6}
		\multicolumn{1}{c|}{}  & \multicolumn{4}{c|}{binary} & dec\\
		\cline{1-6}
		$\overline{x}_0\overline{x}_1$ & $1$ & $1$ & $0$ & $0$ & $12$\\
		$\overline{x}_1$    & $1$ & $1$ & $0$ & $1$ & $13$ \\
		\cline{1-6}
		\multicolumn{1}{c|}{} & $x_3$ & $x_2$ & $x_1$ & $x_0$ & \multicolumn{1}{c}{}\\
		\cline{2-5}
	\end{tabular}
	\hspace{6ex}
	\begin{tabular}{|r|c|c|c|c|r|}
		\cline{2-6}
		\multicolumn{1}{c|}{} & \multicolumn{5}{c|}{index}\\
		\cline{2-6}
		\multicolumn{1}{c|}{}  & \multicolumn{4}{c|}{binary} & dec\\
		\cline{1-6}
		$\overline{x}_2\overline{x}_0$ & $1$ & $0$ & $1$ & $0$ & $10$\\
		$\overline{x}_2$    & $1$ & $0$ & $1$ & $1$ & $11$ \\
		\cline{1-6}
		\multicolumn{1}{c|}{} & $x_3$ & $x_2$ & $x_1$ & $x_0$ & \multicolumn{1}{c}{}\\
		\cline{2-5}
	\end{tabular}
\end{table}

\subsection{Affine transformations as automorphisms}
\label{subsec:autonotLTA}

In the following, we analyze the automorphisms that can be expressed as affine transformations. 
Even though it is not clear if affine transformations include all automorphisms of a polar code (we will provide a counter-example later on), this is a useful starting point to generate UTL automorphisms.  
\begin{theorem}
\label{theo:gen_mon}
An affine transformation belongs to the automorphism group of a given monomial code if and only if it maps all generating monomials into linear combinations of generating monomials. 
\begin{proof}
This property follows from the definition of automorphism of a code. 
In fact, a permutation belongs to the automorphism group of a code if and only if it maps every codeword into another unique codeword, namely if and only if the transformed code book is equal to the original one. 
Codewords of a monomial code correspond to linear combinations of the generating monomials, while affine transformations correspond to permutations.  
\end{proof}
\end{theorem} 
We link this property to the affine transformation matrix $A$.
Non-zero entries of $A$ represent variable changes: setting $A_{i,j}=1$ means to substitute variable $\overline{x}_i$ with variable $\overline{x}_j$. 
Theorem~\ref{theo:gen_mon2} proves that entry $A_{i,j}$ can be set to $1$ only if the variable change does not impact the generating monomial set. 
\begin{theorem}
\label{theo:gen_mon2}
A linear transformation $x'= A x$ with matrix $A$ belongs to the automorphism group of the monomial code generated by monomial set $\mathcal{M}$ if and only if for every $0\leq i,j < n$ we have that $\mathcal{M}_{i,j} \subseteq \mathcal{M}$, where $\mathcal{M}_{i,j} = \mathcal{M}|_{x_i:x_i=A_{i,j}\cdot x_j}$ represents the set of monomials in $\mathcal{M}$ including variable $x_i$ where this variable is substituted by $A_{i,j}\cdot x_j$.
\begin{proof}
If $A_{i,j}=0$, the result of this transformation is the empty set, and hence $\mathcal{M}_{i,j} \subseteq \mathcal{M}$. 
On the other hand, if $A_{i,j}=1$ then $\mathcal{M}_{i,j}$ represents the monomial set generated by the monomials in $\mathcal{M}$ including variable $x_i$, where this variable is substituted by variable $x_j$. 
If we call $\mathcal{M}'$ the set of the generating monomials transformed through $A$, we know that $|\mathcal{M}'| = |\mathcal{M}|$, and hence these two sets include the same number of monomials, since $A$ is a linear bijective transformation. 
Moreover, by construction we have that $\mathcal{M}'$ is composed of linear combinations of elements of $\mathcal{M}_{i,j}$ (with the inclusion of the unity if $1 \in \mathcal{M}$). 
As a consequence, the fact that $\mathcal{M}_{i,j} \subseteq \mathcal{M}$ means that $\mathcal{M}'$ is composed of linear combinations of the elements of subsets of $\mathcal{M}$; this and the fact that $|\mathcal{M}'| = |\mathcal{M}|$ can be true if and only if $\mathcal{M}' = \mathcal{M}$, and Theorem~\ref{theo:gen_mon} concludes the proof. 
\end{proof}
\end{theorem}

As an example, we consider polar code $(16,7)$ generated by the set of monomials $\mathcal{M}=\{1,\overline{x}_0,\overline{x}_1,\overline{x}_2,\overline{x}_0\overline{x}_1,\overline{x}_0\overline{x}_2,\overline{x}_3\}$. 
Entry $A_{1,2}$ can be put to 1; in fact, if we substitute $\overline{x}_1$ by $\overline{x}_2$ in $\mathcal{M}$, we have the set of monomials $\mathcal{M}_{1,2}=\{\overline{x}_2,\overline{x}_0\overline{x}_2,\} \subset \mathcal{M}$. 
On the contrary, if we look at $A_{0,1}$, we have that $\mathcal{M}_{0,1}=\{\overline{x}_1,\overline{x}_1\overline{x}_2\}\not\subset \mathcal{M}$. 
Similarly, for $A_{0,2}$ we have $\mathcal{M}_{0,2}=\{\overline{x}_2,\overline{x}_1\overline{x}_2\}\not\subset \mathcal{M}$. 
Finally, if we substitute any monomial by $x_3$, $\mathcal{M}_{i,3}\not\subset \mathcal{M}$. 
At the end, the only entry above the diagonal that may be put to 1 is $A_{1,2}$. 
A subset of the automorphism group of the code is then given by the affine transformation \eqref{eq:GA} with invertible $A$ and $b$ of the structure
\begin{equation}
\begin{split}
 A=\begin{bmatrix}1&0&0&0\\\star&\star&\star&0\\\star&\star&\star&0\\\star&\star&\star&1\end{bmatrix}
 \end{split}
 , \quad
 \begin{split}
  b= \begin{bmatrix}\star\\\star\\\star\\\star\end{bmatrix} ,
\end{split}
\label{eq:A16_7}
\end{equation}  
increasing the number of automorphisms from 1024 to 2688, however including only one non-trivial UTL transformation.

This procedure can be implemented efficiently using the binary representation of indices in the information set $\mathcal{I}$. 
Given a row of the transformation matrix $T_N$, the variables included in the monomial generating that row are given by the zeroes of the binary expansion of the row index, see Table~\ref{tab:monomials}. 
The impact of an entry $A_{i,j} = 1$ can be checked by tracking the changes in the binary expansions of the information bit indices. 
All indices in $\mathcal{I}$ having $x_i=0$ in their binary expansions are extracted;  then in these indices, $x_i$ is set to 1 while $x_j$ is set to 0, which represents the change of variables.  
If the modified indices are still in $\mathcal{I}$, then $\mathcal{M}_{i,j} \subseteq \mathcal{M}$ and $A_{i,j} = 1$ is feasible. 

We illustrate this method by continuing the previous example, which has the information set $\mathcal{I}=\{7,10,11,12,13,14,15\}$. 
%
For position $A_{1,2}$, we have the original monomials and their extensions in Table~\ref{tab:variable-change} on the left, and those after variable change $x_1 \rightarrow x_2$ on the right.

\subsection{Automorphisms of given polar codes}
\label{subsec:analysisauto}


\begin{figure}[tb]
	\centering
	\includegraphics[width=0.50\textwidth]{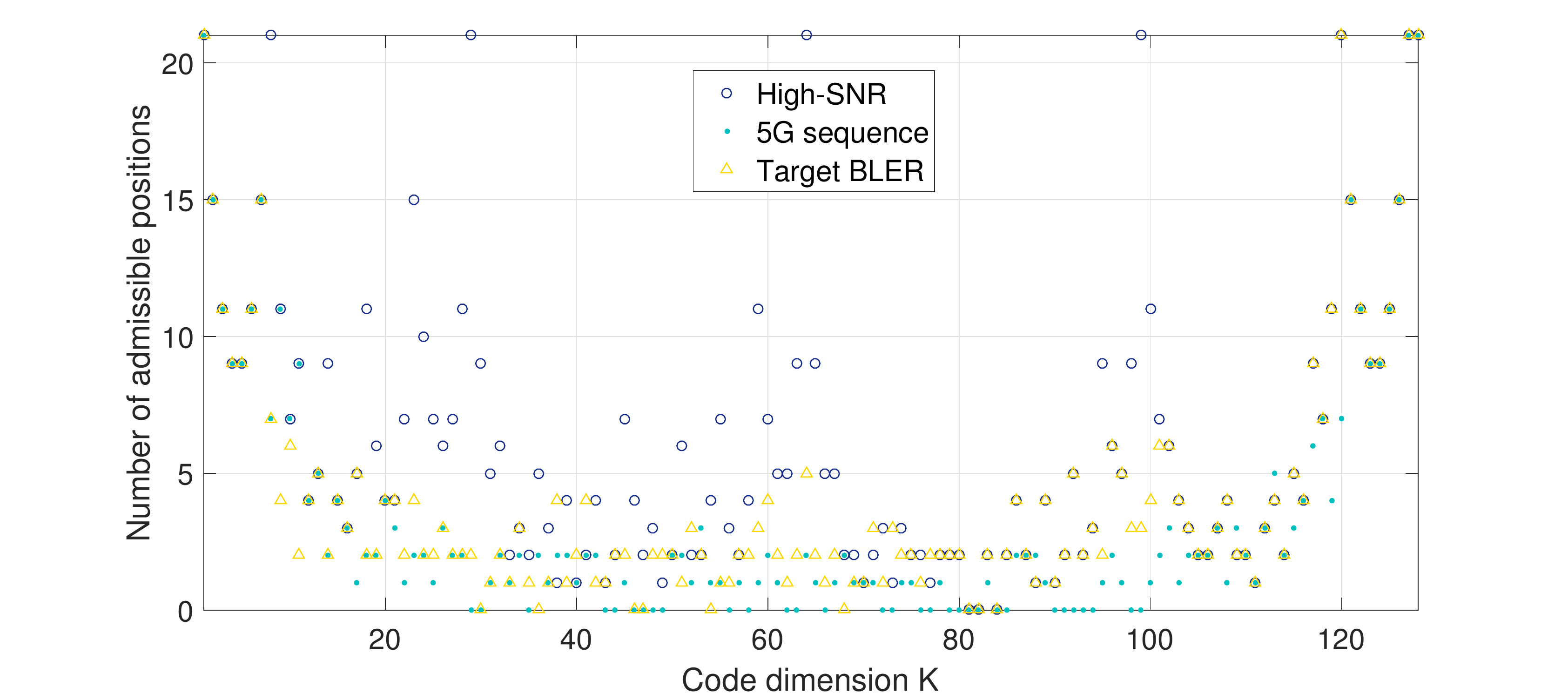}
	\caption{Number of upper-triangular admissible positions in matrix $A$ for polar codes of length $N=128$ and dimension $K$; three different designs.}
	\label{fig:posA-vs-K}
\end{figure} 

Here we analyse three polar code constructions with respect to their UTL automorphisms, namely the 5G sequence \cite{5G,5GHuawei}, a high-SNR DE/GA design at $10.5$~dB, and a low-SNR DE/GA design for the target block error rate (BLER) of $0.001$.  

\begin{table}[t]
	\centering
	\scriptsize
	\caption{Relative number of codes for given length $N$ that have 0 or at least 32 UTL automorphisms; two different designs.}
	\label{tab:num-code-auto-0-32}
	\begin{tabular}{c|c||c|c|c|c}		
		\multicolumn{2}{c||}{} & 128 & 256 & 512 & 1024\\
		\hline
		\hline
		\multirow{2}{*}{High-SNR} & 0 & 0.0234 & 0.0508 & 0.0781 & 0.1396 \\
		\cline{2-2}
		&$\geq32$ & 0.4844 & 0.4414 & 0.3945 & 0.3652\\		
		\hline
		\hline
		\multirow{2}{*}{5G} & 0 & 0.2422 & 0.5742 & 0.7773 & 0.8828 \\
		\cline{2-2}
		&$\geq32$ & 0.2422 & 0.1133 & 0.0664 & 0.0381\\
	\end{tabular}
\end{table}

For a fixed code length $N$ and each code dimension $K$, we determine the number of positions in the upper-triangular (UT) part of matrix $A$ that may be used for automorphisms, termed UT admissible positions, using the methods described above.  The results are shown in Fig.~\ref{fig:posA-vs-K}.  The high-SNR design leads to a larger number of admissible positions; the maximal number of $21$ is attained by the eight Reed-Muller codes.  For the 5G sequence, about one quarter of the codes have zero admissible UT positions.  Note that the considered UTL automorphisms, see \eqref{eq:UTL}, assume $U$ with full diagonal, and thus $t$ UT admissible positions give rise to $2^t$ UTL automorphisms.

An important question is if a code has UTL automorphisms at all and if it has sufficiently many.  For the code lengths $N=128,256,512,1024$ and for each dimension $K$, we determined if the code has $0$ UTL automorphisms, which makes it not suitable for AE decoding, or at least $32$, which makes it suitable.  The relative number of codes, for each code length, fulfilling these criteria is reported in Table~\ref{tab:num-code-auto-0-32} for the high-SNR design and for the 5G sequence.  The number of codes suitable for AE decoding decreases with growing code length, particularly for the 5G sequence.

Finally, we demonstrate that not all automorphisms can be expressed as an affine transformation counting the number of elements of the automorphism group of all polar codes of length $N=8$ by brute force search; the result is reported in Table~\ref{tab:autoPCcount}.
Here we can see that for most polar codes the number of automorphisms, $|\Aut|$, is larger than the number of affine transformations, $|\Aff|$. 
This shows that some automorphisms cannot be expressed as affine transformations.
As an example, consider the $(8,3)$ polar code with $\mathcal{I}=\{5,6,7\}$: swapping 4th and 8th bits is an automorphism that cannot be expressed as an affine transformation.

\subsection{UTL design of polar codes}
\label{subsec:UTLdesign}

The number of UTL automorphisms for reliability-based designs of polar codes decreases with the code length, as discussed above.  In the following, we propose a new method for polar code design that aims at increasing the number of UTL automorphisms by modifying the information set of a given polar code.  The method relies on the following corollary:
\begin{cor}
	\label{cor:gen_mon2}
	A UTL transformation $A$ belongs to the automorphism group of the monomial code generated by monomial set $\mathcal{M}$ if and only if $\mathcal{M}_{i,j} \subseteq \mathcal{M}$ for all $0\leq i<j<n-1$ with $A_{i,j}=1$.
	\begin{proof}
		The proof follows from Theorem~\ref{theo:gen_mon2} and from the fact that it is always true that $\mathcal{M}_{i,i} \subseteq \mathcal{M}$, i.e., elements on the diagonal of $A$ need not be checked. 
	\end{proof}
\end{cor}

\begin{table}[t]
	\centering
	\tiny
	\caption{Number of automorphisms of polar codes of length $N=8$.}
	\label{tab:autoPCcount}
	\begin{tabular}{l|c|c|c|c}
		$\mathcal{M}$ & $K$ & $|\Aut|$ & $|\Aff|$ &   \\
		\hline
		$1$ & 1 & 40,320 & 1344 & $\RM(0,3)$\\
		$1,\overline{x}_0$ & 2 & 1152 & 192 & \\
		$1,\overline{x}_0,\overline{x}_1$ & 3 & 384 & 192 & \\
		$1,\overline{x}_0,\overline{x}_1,\overline{x}_2$ & 4 & 1344 & 1344 & $\RM(1,3)$\\
		$1,\overline{x}_0,\overline{x}_1,\overline{x}_0\overline{x}_1$ & 4 & 384 & 192 & \\
		$1,\overline{x}_0,\overline{x}_1,\overline{x}_0\overline{x}_1,\overline{x}_2$ & 5 & 384 & 192 & \\
		$1,\overline{x}_0,\overline{x}_1,\overline{x}_0\overline{x}_1,\overline{x}_2,\overline{x}_0\overline{x}_2$ & 6 & 1152 & 192 & \\
		$1,\overline{x}_0,\overline{x}_1,\overline{x}_0\overline{x}_1,\overline{x}_2,\overline{x}_0\overline{x}_2,\overline{x}_1\overline{x}_2$ & 7 & 40,320 & 1344 & $\RM(2,3)$
	\end{tabular}
\end{table}

For a code of length $N=2^n$, assume we have the reliability sequence $\mathcal{R}$, i.e., the sequence of bit-channel indices in decreasing reliability order. 
For a code of dimension $K$, the first $K$ entries of $\mathcal{R}$ are used for the information set $\mathcal{I}$.  
To increase the number of UTL automorphisms, the information set is modified with the following method.

We choose an integer $s<K$ and select as reduced information set $\mathcal{I}_s$ the first $K-s$ entries of $\mathcal{R}$; the corresponding generating monomial set is denoted by $\mathcal{M}_s$.  
For a chosen targeted admissible entry $A_{i,j}$ in the upper diagonal part of $A$, the $p\leq s$ monomials are determined that are generated by $A_{i,j}=1$ and not in $\mathcal{M}_s$ yet; this may efficiently be done using the methods from the previous section.  
Adding these $p$ new monomials to $\mathcal{M}_s$ gives a new polar code of dimension $K-s'$, $s'=s-p$, with the new monomial set $\mathcal{M}_{s'}$ and the new information set $\mathcal{I}_{s'}$ according to $\mathcal{M}_{s'}$.
This process is repeated until the desired dimension $K$ is reached; the resulting polar code with information set $\mathcal{I}_0$ has more UT admissible entries in $A$ than the original code. 

As an example, we revisit the $(16,7)$ polar code presented in Section~\ref{subsec:autonotLTA}, which has one non-trivial element of the $\UTL(4)$ subgroup in its automorphism group, see \eqref{eq:A16_7}. 
Choosing $s=1$, we obtain the $(16,6)$ polar code defined by $\mathcal{I}_1=\{7,11,12,13,14,15\}$ or by $\mathcal{M}_1=\{1,x_0,x_1,x_2,x_3,x_0x_1\}$. 
Terming $A_{\mathcal{M}_s}$ the matrix of dimension $n\times n$ having at position $(i,j)$ the monomial(s) needed to be picked to make $A_{i,j}$ part of the automorphism group of the resulting polar codes, then 
\begin{equation}
\scriptsize
	\begin{split}
	A_{\mathcal{M}_1} = \begin{bmatrix}
	[\,] & [\,] & [x_1x_2] & [x_1x_3]\\ 
	[\,] & [\,] & [x_0x_2] & [x_0x_3]\\ 
	[\,] & [\,] &[\,] & [\,] \\ 
	[\,] & [\,] &[\,] & [\,]
	\end{bmatrix}
	\end{split}
	\label{eq:A16_6}
\end{equation}
We retrieve the original $(16,7)$ polar code by including monomial $x_0x_2$, adding the bit-channel index $10$ in $\mathcal{I}_1$. 
If the monomial $x_0x_3$, representing bit channel $6$, is included in $\mathcal{I}_1$, UTL transformations of the form $A^{(6)}$ are automorphisms of the new polar code of dimension $K=7$. 
If monomials $x_1x_2$ (bit channel $9$) or $x_1x_3$ (bit channel $5$) are selected instead, the two $(16,7)$ polar have the UTL transformation matrices $A^{(9)}$ and $A^{(5)}$, allowing either $4$ and $8$ automorphisms from $\UTL(4)$, respectively:
%
\begin{align}
	\tiny
		A^{(6)} &= \left[\begin{smallmatrix}
			1 & 0 & 0 & 0\\ 
			0 & 1 & 0 & \star\\ 
			0 & 0 & 1 & \star \\ 
			0 & 0 & 0& 1
		\end{smallmatrix}\right]
		&
		A^{(9)} &= \left[\begin{smallmatrix}
			1 & \star & \star & 0\\ 
			0 & 1 & 0 & 0\\ 
			0 & 0 & 1 & 0 \\ 
			0 & 0 & 0& 1
		\end{smallmatrix}\right]
		&
		A^{(5)} &= \left[\begin{smallmatrix}
			1 & \star & 0 & \star\\ 
			0 & 1 & 0 & 0\\ 
			0 & 0 & 1 & \star \\ 
			0 & 0 & 0& 1
		\end{smallmatrix}\right]
	\label{eq:A16_7_new}
\end{align}

According to our experience, entries in the lower right corner, i.e. $U_{i,j}$ with $n/2<i<j$, are particularly useful.

\section{Simulation Results}
\label{sec:num}
In this section we present simulation results of polar codes transmitted with BPSK modulation over the AWGN channel.  
As performance reference we use (a) polar codes under SC decoding, where the code is designed for each SNR value by DE/GA (Optimal SNR), and (b) the 5G polar codes \cite{5G,5GHuawei} under SCL and SCL-CRC decoding. 
In the figures, SCL-$L$ refers to SCL decoding with list size $L$; AE$M$-SC, AE$M$-SCL-$L$, AE$M$-BP refers to AE decoding with $M$ branches using SC, SCL-$L$ and BP decoding, respectively.

\begin{figure}[t]
	\includegraphics[width=0.995\columnwidth]{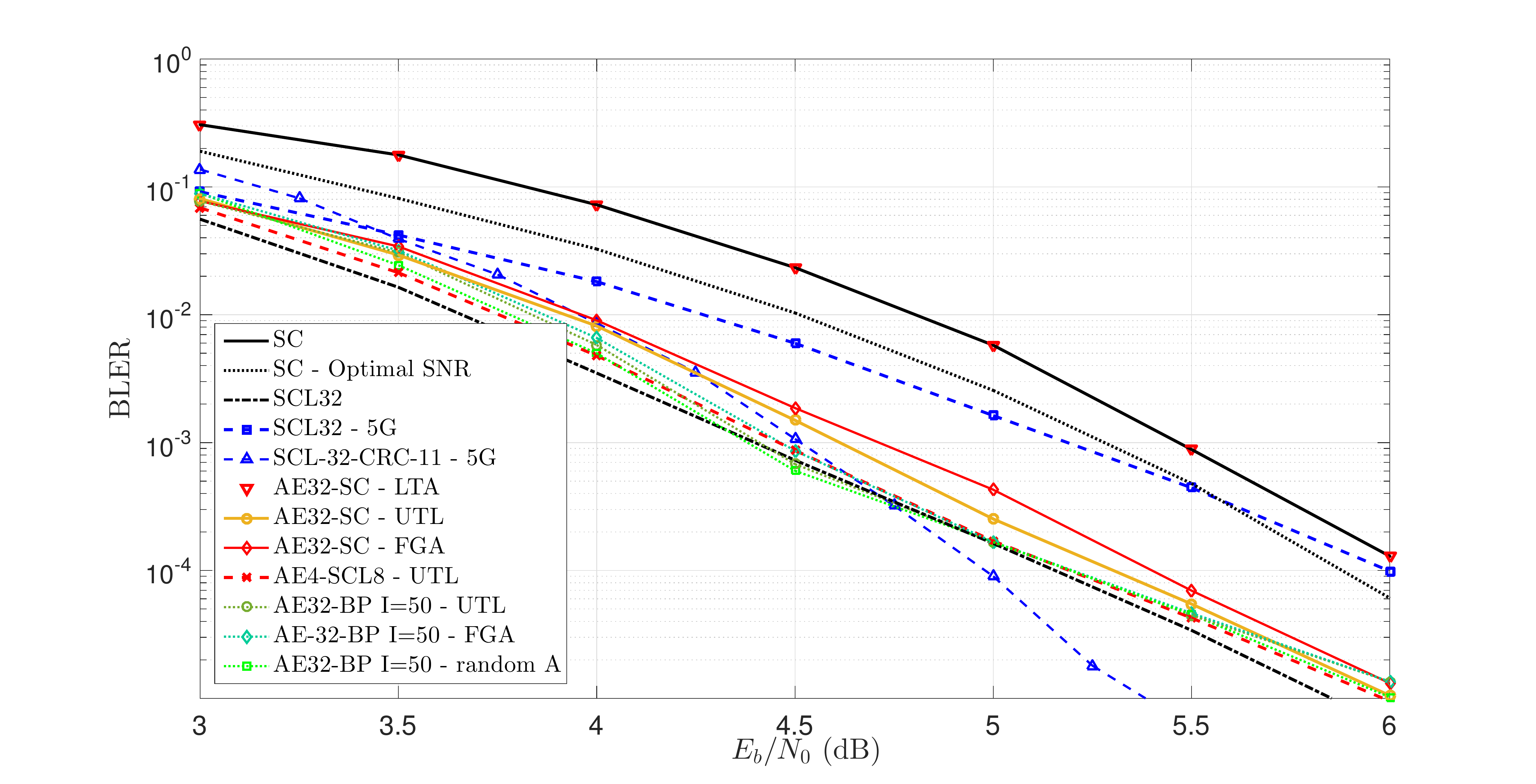}
	\caption{Performance of $(128,100)$ polar codes designed for high SNR.}
	\label{fig:128_100}
\end{figure}

First we consider short polar codes designed using DE/GA for high SNR, i.e. allowing naturally many UTL automorphisms (Table~\ref{tab:num-code-auto-0-32}); the results are shown in Fig.~\ref{fig:128_100}.
For AE decoding, the labels indicate which automorphisms are used. 
As expected from theory \cite{geiselhart2020automorphism}, LTA automorphisms provide no gain for AE-SC decoding.  
Moreover, current state-of-the-art permuted decoder was using factor graph permutations, with a priority on factor graph automorphisms (FGA) \cite{PermDecRussian} that do not degrade the error-capability of the code. Here we show that UTL automorphisms improves permuted decoding compared to FGA and provide performance close to SCL decoding while keeping SC latency. 
Performance of AE-BP is even better and is moreover improved by using larger automorphism groups since LTA is not absorbed for BP decoding.

\begin{figure}[t!]
	\includegraphics[width=0.995\columnwidth]{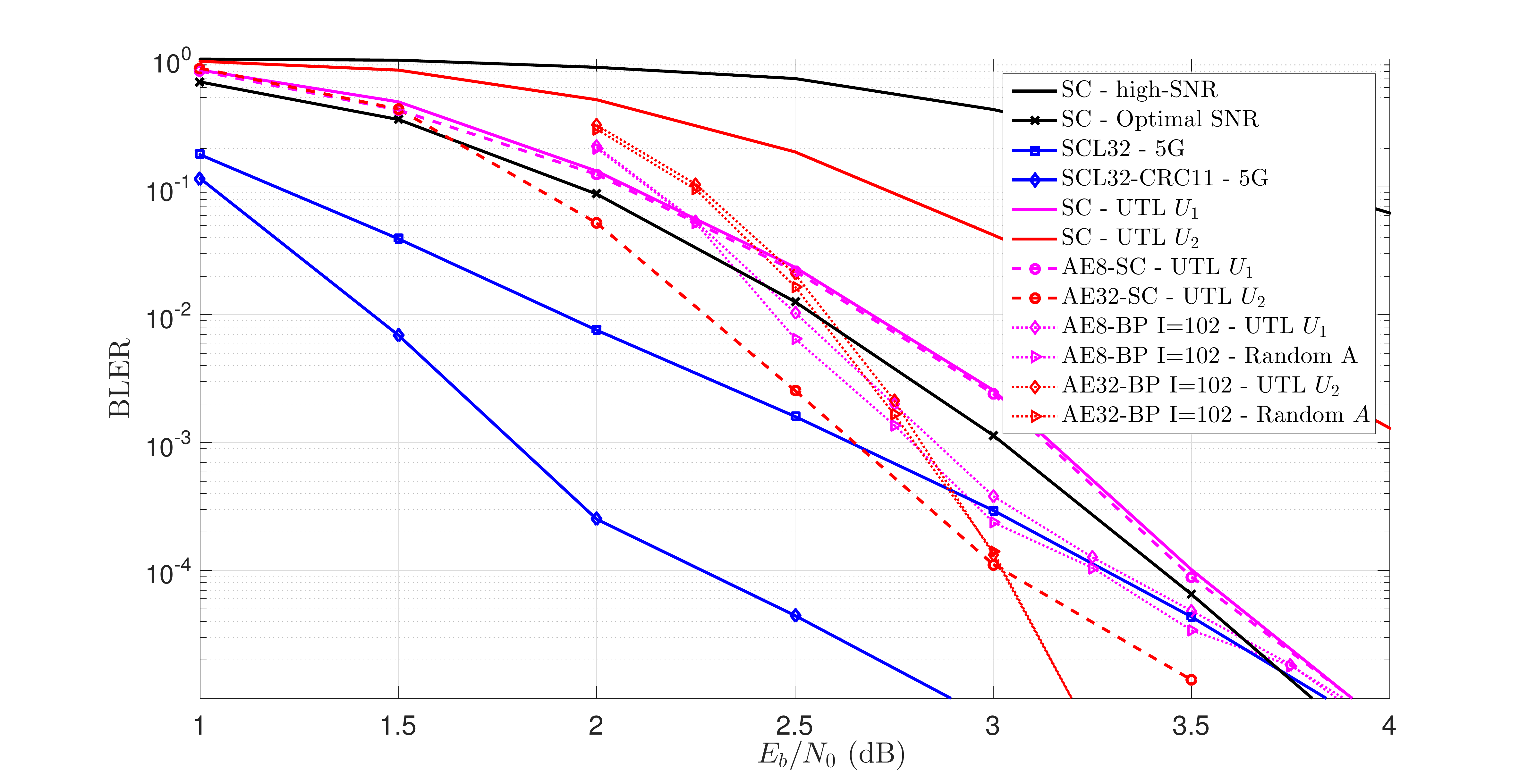}
	\caption{Performance of $(1024,512)$ polar codes by UTL design.}
	\label{fig:1024_512}
\end{figure}

Second we consider longer polar codes, for which design based on reliability leads only to a small number of UTL automorphisms if to any at all, and we apply our proposed UTL design to ensure sufficiently many UTL automorphisms.
The results are shown in Fig.~\ref{fig:1024_512}.
The polar code has length $N = 1024$ and dimension $K=512$.  Note that this dimension is far from the closest possible Reed-Muller code dimensions, which are $386$ and 638, leading to a small number of UTL automorphisms; for the high-SNR design there are only $2$ UTL automorphisms. 
The loss inherent to a inaccurate design SNR grows with the code length, which is confirmed by the $1.75$~dB loss of this high-SNR code under SC decoding as compared to the code with optimal SNR design.
%
We provide two UTL designs, corresponding to transformation matrices 
%
%
\begin{align}
	U_1 &= \left[\begin{smallmatrix}
		\vrule width 0pt depth 1pt \\
		1 & \star & \star & 0 & 0 & 0  & 0 & 0 & 0 & 0\\ 
		0 & 1 & \star & 0 & 0 & 0  & 0 & 0 & 0 & 0\\ 
		0 & 0 & 1& 0 & 0 & 0  & 0 & 0 & 0 & 0\\ 
		0 & 0 & 0& 1 & 0 & 0  & 0 & 0 & 0 & 0 \\
		0 & 0 & 0& 0 & 1 & 0  & 0 & 0 & 0 & 0 \\ 
		0 & 0 & 0& 0 & 0 & 1  & 0 & 0 & 0 & 0 \\ 
		0 & 0 & 0& 0 & 0 & 0  & 1 & 0 & 0 & 0 \\ 
		0 & 0 & 0& 0 & 0 & 0  & 0 & 1 & 0 & 0 \\ 
		0 & 0 & 0& 0 & 0 & 0  & 0 & 0 & 1 & 0 \\ 
		0 & 0 & 0& 0 & 0 & 0  & 0 & 0 & 0 & 1 \\[3pt]
	\end{smallmatrix}\right]
	&
	U_2 &= \left[\begin{smallmatrix}
		\vrule width 0pt depth 1pt \\
		1 & \star & 0 & 0 & 0 & 0  & 0 & 0 & 0 & 0\\ 
		0 & 1 & 0 & 0 & 0 & 0  & 0 & 0 & 0 & 0\\ 
		0 & 0 & 1 & \star  & 0 & 0  & 0 & 0 & 0 & 0\\ 
		0 & 0 & 0& 1  & 0 & 0  & 0 & 0 & 0 & 0 \\
		0 & 0 & 0& 0 & 1 & 0  & 0 & 0 & 0 & 0 \\ 
		0 & 0 & 0& 0 & 0 & 1 & 0 & 0 & 0 & 0 \\ 
		0 & 0 & 0& 0 & 0 & 0  & 1 & 0 & 0 & 0 \\ 
		0 & 0 & 0& 0 & 0 & 0  & 0 & 1 & \star & \star \\ 
		0 & 0 & 0& 0 & 0 & 0  & 0 & 0 & 1 & \star \\ 
		0 & 0 & 0& 0 & 0 & 0  & 0 & 0 & 0 & 1  \\[3pt]
	\end{smallmatrix}\right]
	\label{eq:A1024_512}
\end{align}
in order to demonstrate the impact of the positions of the ones in the UTL matrix (UT admissible positions) on the AE decoder. 
The first UTL design focuses on the upper left part of the UTL matrix.  
The polar code is designed by human inspection starting from $s=11$, and it has $8$ UTL automorphisms, represented by $U_1$. 
For the second polar code we aim for ones in the lower right corner of the UTL matrix, starting with $s=60$.  
Simulations show that the second code has a lower error rate than the first, and we conjecture that this is due to the UTL automorphisms corresponding to 1s in the lower right corner. 
This corner was as well targeted in factor graph permutation\cite{PermGross} and is actually crucial when SC is used. 
Whereas AE-BP provides already more diversity even with the first configuration. However, the lower-block improves even more performance of AE-BP.
Further, the second code is better than the 5G polar code under SCL decoding, though still outperformed by the 5G polar code under SCL-CRC decoding.


%
\section{Conclusions}
\label{sec:conclusions}
In this paper we analyze automorphisms of polar codes, specifically automorphisms of the upper-triangular linear (UTL) group.  
While for short polar codes a conventional high-SNR design using DE/GA already leads to many UTL automorphisms, longer codes require specific designs.  
We propose a method to design polar codes that have sufficiently many UTL automorphisms to take advantage of automorphism ensemble (AE) decoding.
According to our analysis, the bottom right-part of the UTL matrix should be privileged in order to get a polar code well-designed for AE decoding.  

These preliminary results show that our approach is very promising. Still, many questions remain open for future research.
First, the proposed design principle is guided by human inspection, and an algorithm to fully automatize the code design is still under research.
Then, it is not clear how to design UTL matrices to improve for AE decoding, and why the bottom right part of these matrices seems important. 
Finally, additional CRC seems not to improve AE decoding, making SCL-CRC decoding still preferable when no constraints on decoding latency are given. 

\balance
\bibliographystyle{IEEEbib}
\bibliography{references}

\begin{thebibliography}{10}

\bibitem{ArikanFirst}
E.~Arikan,
\newblock ``Channel polarization: a method for constructing capacity-achieving
  codes for symmetric binary-input memoryless channels,''
\newblock {\em IEEE Transactions on Information Theory}, vol. 55, no. 7, pp.
  3051--3073, July 2009.

\bibitem{TalSCL}
I.~Tal and A.~Vardy,
\newblock ``List decoding of polar codes,''
\newblock {\em IEEE Transactions on Information Theory}, vol. 61, no. 5, pp.
  2213--2226, May 2015.

\bibitem{CRC_aid}
K.~Niu and K.~Chen,
\newblock ``{CRC}-aided decoding of polar codes,''
\newblock {\em IEEE Communications Letters}, vol. 16, no. 10, pp. 1668--1671,
  October 2012.

\bibitem{PermGross}
N.~Doan, S.~A. Hashemi, M.~Mondelli, and W.~J. Gross,
\newblock ``On the decoding of polar codes on permuted factor graphs,''
\newblock in {\em IEEE Global Communications Conference (GLOBECOM)}, Abu Dhabi,
  UAE, Dec. 2018.

\bibitem{SCANfirst}
U.~U. Fayyaz and J.~R. Barry,
\newblock ``Low-complexity soft-output decoding of polar codes,''
\newblock {\em IEEE Journal on Selected Areas in Communications}, vol. 32, no.
  5, pp. 958--966, May 2014.

\bibitem{SCANL}
C.~Pillet, C.~Condo, and V.~Bioglio,
\newblock ``{SCAN} list decoding of polar codes,''
\newblock in {\em IEEE International Conference on Communications (ICC)},
  Dublin, Ireland, June 2020.

\bibitem{BPfirst}
E.~Arikan,
\newblock ``Polar codes : a pipelined implementation,''
\newblock in {\em International Symposium on Broadband Communications (ISBC)},
  Melaka, Malaysia, July 2010.

\bibitem{BPLRM}
A.~Elkelesh, M.~Ebada, S.~Cammerer, and S.~ten Brink,
\newblock ``Belief propagation list decoding of polar codes,''
\newblock {\em IEEE Communications Letters}, vol. 22, no. 8, pp. 1536--1539,
  Aug 2018.

\bibitem{RMcode}
D.~E. Muller,
\newblock ``Application of boolean algebra to switching circuit design and to
  error detection,''
\newblock {\em Transactions of the I.R.E. Professional Group on Electronic
  Computers}, vol. EC-3, no. 3, pp. 6--12, 1954.

\bibitem{RMcodeREED}
I.~Reed,
\newblock ``A class of multiple-error-correcting codes and the decoding
  scheme,''
\newblock {\em Transactions of the IRE Professional Group on Information
  Theory}, vol. 4, no. 4, pp. 38--49, 1954.

\bibitem{geiselhart2020automorphism}
M.~Geiselhart, A.~Elkelesh, M.~Ebada, S.~Cammerer, and S.~Ten~Brink,
\newblock ``Automorphism ensemble decoding of {Reed-Muller} codes,''
\newblock {\em IEEE Transactions on Communications}, pp. 1--1, 2021.

\bibitem{BardetPolyPC}
M.~Bardet, V.~Dragoi, A.~Otmani, and J.~Tillich,
\newblock ``Algebraic properties of polar codes from a new polynomial
  formalism,''
\newblock in {\em 2016 IEEE International Symposium on Information Theory
  (ISIT)}, Barcelona, Spain, July 2016.

\bibitem{Urbanke_symmetry}
Ivanov K. and Urbanke R.,
\newblock ``On the dependency between the code symmetries and the decoding
  efficiency,''
\newblock in {\em arXiv preprint arXiv:2001.03790}, Feb 2021.

\bibitem{geiselhart2021automorphismPC}
M.~Geiselhart, A.~Elkelesh, M.~Ebada, S.~Cammerer, and S.~ten Brink,
\newblock ``On the automorphism group of polar codes,''
\newblock in {\em arXiv preprint arXiv:2101.09679}, Jan 2021.

\bibitem{frozenset}
H.~Vangala, E.~Viterbo, and Y.~Hong,
\newblock ``A comparative study of polar code constructions for the {AWGN}
  channel,''
\newblock in {\em arXiv preprint arXiv:1501.02473}, 2015.

\bibitem{5G}
$3^{\text{rd}}$ Generation Partnership Project~({3GPP}),
\newblock ``Multiplexing and channel coding,''
\newblock {\em 3GPP 38.212 V.15.3.0}, 2018.

\bibitem{5GHuawei}
V.~Bioglio, C.~Condo, and I.~Land,
\newblock ``Design of polar codes in {5G New Radio},''
\newblock {\em IEEE Communications Surveys Tutorials}, vol. 23, no. 1, Mar.
  2021.

\bibitem{PermDecRussian}
M.~Kamenev, Y.~Kameneva, O.~Kurmaev, and A.~Maevskiy,
\newblock ``Permutation decoding ofpolar codes,''
\newblock in {\em XVI International Symposium "Problems of Redundancy in
  Information and Control Systems" (REDUNDANCY)}, Moscow, Russia, Oct. 2019.

\end{thebibliography}

\end{document}